\documentclass[%aip,
jmp,
amsmath,amssymb,
preprint,%
numerical,
]{revtex4-1}
\usepackage{graphicx}% Include figure files
\usepackage{dcolumn}% Align table columns on decimal point
\usepackage{bm}% bold math
\usepackage{hyperref}
\usepackage[utf8]{inputenc}
\usepackage[T1]{fontenc}
\usepackage{mathptmx}
\usepackage{etoolbox}
\newcommand{\RNum}[1]{\uppercase\expandafter{\romannumeral #1\relax}}
\usepackage{array,xcolor,colortbl}
\usepackage{multirow}
\newcommand{\ud}{\ensuremath{\mathrm{d}}}

\newcommand{\bcdot}{\ensuremath{\boldsymbol\cdot}}

\newcommand{\beq}{\begin{equation}}
\newcommand{\eeq}{\end{equation}}

\makeatletter
\def\@email#1#2{%
	\endgroup
	\patchcmd{\titleblock@produce}
	{\frontmatter@RRAPformat}
	{\frontmatter@RRAPformat{\produce@RRAP{*#1\href{mailto:#2}{#2}}}\frontmatter@RRAPformat}
	{}{}
}%
\makeatother

\begin{document}
	
\title{Phototactic bioconvection under oblique collimated irradiation in a forward scattering suspension}
% Force line breaks with \\
	
% Force line breaks with \\
\author{A. Rathi}
\altaffiliation[Corresponding author: E-mail: ]{rathi.iiitdmj@gmail.com.}%Lines break automatically or can be forced with \\
\author{M. K. Panda}
\author{S. K. Rajput}%
%\email{mkpanda@iiitdmj.ac.in}
%\email{shubh.iiitj@gmail.com}
%\email{rathi.iiitdmj@gmail.com}
\affiliation{Department of Mathematics, PDPM Indian Institute of Information Technology Design and Manufacturing, Jabalpur 482005, India.%\\This line break forced with \textbackslash\textbackslash
}
%
%\homepage{http://www.Second.institution.edu/~Charlie.Author.}
%\affiliation{%
%	Second institution and/or address%\\This line break forced% with \\
%}%
	

%%%%%%%%%%%%%%%%%%%%%%%%%%%%%%%%%%%%%%%%%%%%%%%%%%%%%%%%%%%%%%%%%%%%%%%%%%%		
\begin{abstract}
		
Phototaxis, the process by which living organisms navigate toward optimal light conditions, is essential for motile photosynthetic microorganisms. Positive(negative) phototaxis denotes the motion directed towards(away from) the source of illumination. The main objective of this study is the numerical investigation of onset of bioconvection in a suspension of phototactic microorganisms illuminated by oblique collimated irradiation at the top. In this suspension, the algal cells absorb and anisotropically scatter incident light which influences the flow dynamics of the cells.

\end{abstract}
	
%%%%%%%%%%%%%%%%%%%%%%%%%%%%%%%%%%%%%%%%%%%%%%%%%%%%%%%%%%%%%%%%%%%%%%%%%%%	
	
\maketitle
	
%%%%%%%%%%%%%%%%%%%%%%%%%%%%%%%%%%%%%%%%%%%%%%%%%%%%%%%%%%%%%%%%%%%%%%%%%%%	
	
\section{INTRODUCTION}
	
In 1961, Platt first coined the term "bioconvection" to describe the pattern formation observed in shallow suspensions of motile microorganisms which are denser than the surrounding fluid in which they swim~\cite{12platt1961}. Bioconvection refers to the spontaneous formation of patterns in a suspension of motile microorganisms due to their collective movement~\cite{13pedley1992,14hill2005}. These microbes, being slightly denser than the ambient fluid, on average, tend to swim upward direction. When microorganisms, such as motile algae and bacteria, respond to external stimuli, such behaviour known as taxis—they induce density variations that can lead to convective instability. Important types of taxis encompass gravitaxis, gyrotaxis, chemotaxis and phototaxis~\cite{13pedley1992}. Gravitaxis is the motion of microorganisms in responce to  gravity or acceleration, where upward swimming is termed negative gravitaxis(geotaxis). The swimming orientation of microorganisms, particularly those with a bottom-heavy structure, is governed by the balance between hydrodynamic and gravitational torques, a phenomenon known as gyrotaxis~\cite{1kessler1985}. Chemotaxis, a response to chemical gradients by the microorganisms. Phototaxis, on the other hand, refers to the movement of microorganisms in response to light, where positive phototaxis describes motion towards the light source, while negative phototaxis describes movement away from it. In general, each of these factors reorients an individual microorganism in a specific direction, resulting in an average directed motion rather than the random diffuse motion associated with microorganisms. This study focuses exclusively on phototaxis and its role in bioconvection. \\
Bioconvection patterns are usually observed in the laboratory in shallow suspensions~\cite{14hill2005}. Experimental studies have demonstrated that illumination influence and alter the bioconvection patterns~\cite{21wager1911vii,1kessler1985,22vincent1995mathematical}. Light can also affect the shape and size of the bioconvection pattern~\cite{1kessler1985}. The potential factors contributing to changes in bioconvection patterns due to light intensity are given as follows: First, the motile microorganisms are phototactic, they exhibit positive phototaxis, moving toward the light source, when the light intensity \( G \) is below a critical threshold \( G_c \). However, when the intensity exceeds \( G_c \), they display negative phototaxis, swimming away from the light source. Consequently, the algal cells tend to aggregate at optimal locations in their environment where the light intensity approaches the critical value, \( G \approx G_c \)~\cite{3hader1987}. The second factor contributing to the variation in the patterns is the absorption and scattering of light by microorganisms, which causes the light intensity \(G\) to vary along its path of incidence. In a suspension of finite depth illuminated from above, a basic horizontally uniform state exists. This state arises from the balance between phototaxis, influenced by suspension shading, and diffusion caused by the random component of the cell's swimming motion. Resulting a horizontal concentrated layer(sublayer) of microbes is formed, the location of which depends on the critical light intensity. The region above this sublayer is gravitationally stable and rests on the gravitationally unstable region below it. When the entire fluid layer becomes unstable, fluid motion within the unstable region must extend into the overlying stable layer. This phenomenon, known as penetrative convection, is observed in various convection-related problems~\cite{5straughan1993}. \\
Over time, significant progress has been made in understanding the mechanisms of bioconvection. Initial research primarily examined gravitactic microorganisms, analyzing the swimming behavior of dense motile microbes to determine the instability conditions that give rise to convective flow. Since many motile microorganisms rely on photosynthesis, they exhibit strong phototactic behavior, making it essential to incorporate phototaxis into the realistic models of their movement. The significance of phototaxis and the resulting bioconvection is crucial and should not be overlooked in the field of biofluid dynamics research~\cite{2williams2011}. Vincent and Hill~\cite{8vincent1996} were the first to develop a model describing the behavior of a purely phototactic algal suspension exposed to uniform illumination from above. A purely phototactic response implies that the upswimming of cells is driven by phototaxis only, without any contribution from mass anisotropy. Through a linear stability analysis of the equilibrium state, they identified both stationary and oscillatory disturbance modes at the onset of instability. Later, Ghorai and Hill~\cite{6ghorai2005} conducted numerical simulations of two-dimensional phototactic bioconvection in the \( xz \)-plane using the model proposed by Vincent and Hill~\cite{8vincent1996}. In both studies, the effect of light scattering by algal cells within the suspension was not considered. Ghorai \textit{et al.}~\cite{4ghorai2010} was initially examined the effect of scattering on phototactic bioconvection within isotropic-scattering algal suspensions. They also revealed that the emergence of a bimodal steady state is a consequence of light scattering. The scattering of light by algae is influenced by their shape and size parameters. To analyze the impact of forward scattering on the onset of bioconvection, Ghorai and Panda~\cite{23ghorai2013} introduced a mathematical model. Subsequently, Panda and Ghorai~\cite{9panda2013} investigated bioconvective instability in a suspension of microbes that exhibits isotropic light scattering and they simulated numerically the governing bioconvective system for the same suspension in a two dimensional domain. Furthermore, Panda and Singh~\cite{7panda2016} examined how lateral rigid walls affect the onset and stability of phototactic bioconvection in a two-dimensional geometry within the nonlinear regime. To investigate photo-bioconvection under natural environmental conditions, Panda \textit{et al.}~\cite{24panda2016} examined the combined effects of diffuse and collimated solar flux in an isotropic scattering algal suspension. This work was further extended by Panda~\cite{25panda2020} who explored the influence of diffuse and collimated irradiation on the onset of photo-bioconvection in an anisotropic scattering medium. Next, Panda \textit{et al.}~\cite{26panda2022} investigated the effect of an oblique collimated radiative flux on bioconvective instability in a non-scattering algal suspension. Afterward, Panda and Rajput~\cite{27rajput2023} studied the combined effects of diffuse and oblique collimated radiative flux in a uniformly scattering (isotropic) algal suspension.  \\ 
We study phototactic bioconvection in a two-dimensional domain bounded by a rigid lower surface and a stress-free upper surface. The suspension is illuminated from above by an obliquely incident collimated light beam. It is assumed that the algal cells absorb and scatter the incident light and the scattering is considered to be linearly anisotropic in the forward direction.
\\
\\
This paper is organized as follows: \textit{Section.2} presents the mathematical model, including the governing equations and associated boundary conditions. In  \textit{Section.3}, we describe the steady solution and the linear stability analysis used to derive instability criterion. \textit{Section.4} provides numerical results, analyzing the effects of various parameters on bioconvection behavior. Finally, in \textit{Section.5}, we summarize the results from the numerical experiment.

\section{MATHEMATICAL FORMULATION}

We consider an infinite horizontal two-dimensional fluid layer containing a large number of phototactic microorganisms in the \(x\)-\(z\) cartesian plane, where the \(z\)-axis is oriented vertically upwards. Also, this layer is bounded in the vertical direction with depth \(H\). Here, it is assumed that the top wall is stress-free, whereas a no-slip (rigid) boundary condition is applied at the bottom wall. The light intensity is subjected to non-reflecting boundary conditions at the horizontal boundaries.  In this study, an oblique collimated beam of light enters the suspension from above, striking the upper surface at a fixed off-normal angle \( \theta_i \). The angle of refraction \( \theta_0 \), describing the direction of the collimated beam as it travels through the water, is calculated by using Snell's law, i.e., \(\frac{\sin \theta_i}{\sin\theta_0} = n_0\), where \( n_0 \) is the refractive index of water, taken to be approximately 1.333. 

\begin{figure}[!htbp]
	\centering
	\includegraphics[scale=0.45]{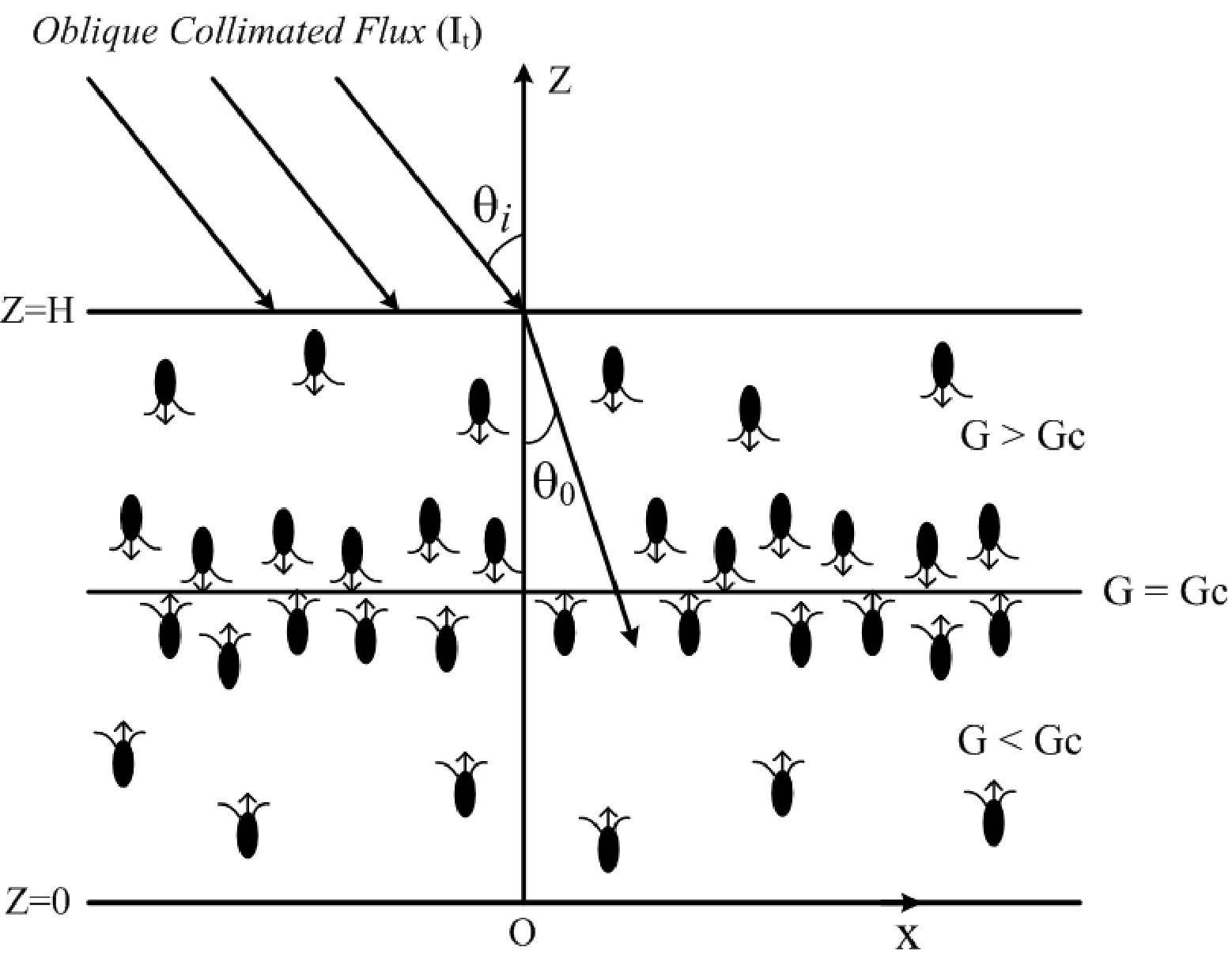}
	\caption{\footnotesize{Formation of the sublayer at \(G=G_c\), where \(G_c\) is the critical light intensity.}}
	\label{fig1}
\end{figure}

\subsection{The Governing Equations}
Consistent with previous models of bioconvection~\cite{13pedley1992}, we consider a monodisperse population of motile microorganisms that can be described by a continuous distribution. The suspension is assumed to be dilute, implying that the volume fraction of cells is small and that cell–cell interactions are negligible. Each microorganism has a volume \( \vartheta \), with a density difference \( \Delta \rho \) relative to the ambient fluid density \( \rho \). Let \( \boldsymbol{u} = u \,\hat{\boldsymbol{x}} + w \,\hat{\boldsymbol{z}} \) and \( n \) represent the average velocity of all the material and cell concentration over an elemental volume. Assuming that the suspension is incompressible and introducing the stream function $\psi$ and vorticity $\zeta$. Thus, the governing equations are given as
\begin{equation}\label{1}
	\boldsymbol{u} = (u,0,w) = \left(\frac{\partial \psi}{\partial z}, 0, \frac{-\partial \psi}{\partial x}\right), \quad \zeta = -\boldsymbol{\nabla}^2 \psi
\end{equation}
	
Neglecting all forces acting on the fluid except for the cell's negative buoyancy, given by \( n \vartheta \Delta \rho g \) per unit volume (where \( g \) is the acceleration due to gravity). Therefore, the momentum equation under the Boussinesq approximation yields	
\begin{align}\label{2}
\frac{\partial \zeta}{\partial t}+\boldsymbol{\nabla}\cdot(\zeta\boldsymbol{u}) = \nu\boldsymbol{\nabla}^2\zeta  - \frac{\Delta \rho g \vartheta}{ \rho}\frac{\partial n}{\partial x}.
\end{align}
Next, the Conservation equation of cells is defined by
\begin{equation}\label{3}
	\frac{\partial n}{\partial t} + \boldsymbol{\nabla}\bcdot \boldsymbol{J} = 0,
\end{equation}
where the total flux $\boldsymbol{J}$ is given by
\begin{equation*}
	\boldsymbol{J} = n \boldsymbol{u} + n \boldsymbol{W_c} - \boldsymbol{D} \boldsymbol{\nabla} n.
\end{equation*}

The cell flux expression comprises three terms: the first term corresponds to the advection of algal cells caused by the bulk fluid flow, while the second term accounts for the mean swimming of the cells. The third term represents the random(diffusive) motion of the cells. This formulation is based on the assumption that the microorganisms are purely phototactic, therefore, the viscous torque effects in horizontal swimming are neglected. Additionally, the diffusion tensor for the cells is assumed to be isotropic and constant, represented as, \( \boldsymbol{D} = D \boldsymbol{I} \), where \(D\) is the cell diffusivity coefficient~\cite{6ghorai2005,9panda2013}. 

%%%%%%%%%%%%%%%%%%%%%%%%%%%%%%%%%%%%%%%%%%%%%%%%%%%%%%%%%%%%%%%%%%%%%%

\subsection{The average swimming orientation}

Let \( I(\boldsymbol{x},\boldsymbol{s}) \) represent the radiation intensity propagating in the unit direction \( \boldsymbol{s} \) at position \( \boldsymbol{x}\) within the suspension. The unit direction vector \( \boldsymbol{s} \) is expressed in spherical polar coordinates and is given by \(\boldsymbol{s} = \sin\theta\cos\phi \,\hat{\boldsymbol{x}} + \cos\theta \,\hat{\boldsymbol{z}}\), where \( \theta \) and \( \phi \) are polar and azimuthal angle. The suspension is exposed to an oblique collimated irradiation from above. To compute the light intensity profiles in a medium that exhibits both absorption and scattering, the radiative transfer equation(hereafter abbreviated as RTE) is employed and is given by
\begin{equation}\label{4}
	\boldsymbol{s} \bcdot \boldsymbol{\nabla} I(\boldsymbol{x},\boldsymbol{s}) + (\sigma_a + \sigma_s)I(\boldsymbol{x},\boldsymbol{s}) = \frac{\sigma_s}{4\pi}\int_{0}^{4\pi}I(\boldsymbol{x},\boldsymbol{s}') \Lambda(\boldsymbol{s}',\boldsymbol{s}) d\Omega'
\end{equation}  
where \(\sigma_a\) is the extinction coefficient, \(\sigma_s\) is the scattering coefficient, and \(\Omega'\) is the solid angle. The scattering phase function, \(\Lambda(\boldsymbol{s}',\boldsymbol{s})\),
is the probability density function for scattering from direction \(\boldsymbol{s}'\) to direction \(\boldsymbol{s}\). For simplicity, we assume \(\Lambda(\boldsymbol{s}',\boldsymbol{s})\) to be linearly anisotropic with azimuthal symmetry
\begin{equation}\label{5}
	\Lambda(\boldsymbol{s}',\boldsymbol{s}) = 1 + A \cos\theta \cos\theta'
\end{equation}
where \(\theta\) and \(\theta'\) are the polar angles from the \(z\)-axis describing the unit vector \(\boldsymbol{s}\) and \(\boldsymbol{s}'\) respectively, and \(A\in[-1,1]\) is the linearly anisotropic scattering coefficient.
\begin{equation*}
	A < 0 : \text{Backward Scattering}, \quad
	A > 0 : \text{Forward Scattering}, \quad
	A = 0 : \text{Isotropic Scattering}.
\end{equation*}

The intensity at the top boundary surface at the location \(\boldsymbol{x_b} = (x,z=H)\) is 
\begin{equation}\label{6}
	I(\boldsymbol{x_b},\boldsymbol{s}) = I_t \delta (\boldsymbol{s}-\boldsymbol{s}_0)
\end{equation}
where \(I_t\) is the magnitude of the incident oblique collimated radiation. Also \(\boldsymbol{s}_0 = \sin(\pi-\theta_0) \cos \phi_0 \,\hat{\boldsymbol{x}} + \cos(\pi-\theta_0) \,\hat{\boldsymbol{z}}\) is the incident direction, where \(\hat{\boldsymbol{x}},\hat{\boldsymbol{z}}\) are unit vectors along the \(x,z\) axes, respectively. The Dirac-delta function, \(\delta\), satisfies
\begin{equation}\label{7}
	\int_{0}^{4\pi} f(\boldsymbol{s}) \delta (\boldsymbol{s}-\boldsymbol{s}_0) d \Omega = f(\boldsymbol{s}_0)
\end{equation}

The absorption and scattering both are considered to be linearly proportional to the concentration of algal cells so that \(\sigma_a = \kappa n(\boldsymbol{x})\) and \(\sigma_s = \varsigma n(\boldsymbol{x})\). In terms of the single scattering albedo \(\omega = \sigma_s/(\sigma_a + \sigma_s)\), the RTE can be written as
\begin{equation}\label{8}
	\boldsymbol{s} \bcdot \boldsymbol{\nabla} I(\boldsymbol{x},\boldsymbol{s}) + \beta I(\boldsymbol{x},\boldsymbol{s}) = \frac{\omega \beta}{4\pi}\int_{0}^{4\pi}I(\boldsymbol{x},\boldsymbol{s}') (1 + A \cos\theta \cos\theta') d\Omega'
\end{equation}
where \(\beta = (\kappa+\varsigma)n\) is the extinction coefficient. Also, \(\omega \in [0,1]\) and \(\omega=0\) for a purely absorbing medium whereas \(\omega=1\) represents a purely scattering medium.
The total intensity \(G(\boldsymbol{x})\) and the radiative heat flux \(\boldsymbol{q(x)}\) at a point \(\boldsymbol{x}\) in the suspension are  
\begin{equation}\label{9}
	G(\boldsymbol{x}) = \int_{0}^{4\pi} I(\boldsymbol{x},\boldsymbol{s}) d \Omega \quad \text{and} \quad \boldsymbol{q(x)} = \int_{0}^{4\pi} I(\boldsymbol{x},\boldsymbol{s}) \boldsymbol{s} d \Omega.
\end{equation} 
For many species of microorganisms, the swimming speed remains independent of illumination, position, time, and direction~\cite{10hill1997}. Let \( W_c \) be the average swimming speed of algal cells. Therefore, the mean swimming velocity \(\boldsymbol{W_c}\) is defined as
\begin{equation}\label{10}
	\boldsymbol{W_c} = W_c \langle \boldsymbol{p} \rangle.
\end{equation}
Here, the average swimming direction, \( \langle \boldsymbol{p} \rangle \), is formulated as
\begin{equation}\label{11}
	\langle \boldsymbol{p} \rangle = -T(G) \frac{\boldsymbol{q}}{|\boldsymbol{q}|},
\end{equation}
where \( T(G) \) is the phototaxis function is given by
\begin{equation}\label{12}
	T(G) =
	\begin{cases} 
		\geq 0, & G(\boldsymbol{x}) \leq G_c, \\
		< 0, & G(\boldsymbol{x}) > G_c.
	\end{cases}
\end{equation}
Here, \(T(G)\) is a function of light intensity and its functional form is depend on the species of microorganisms. The negative sign in Eq.~(\ref{11}) reflects the fact that the light source, from the perspective of a microorganism, is located in the opposite direction to the radiative heat flux vector.

%%%%%%%%%%%%%%%%%%%%%%%%%%%%%%%%%%%%%%%%%%%%%%%%%%%%%%%%%%%%%%%%%%%%%%%%

\subsection{Boundary Conditions}
We impose rigid, no-slip boundary condition at $z = 0$, and require that both the normal velocity and tangential stress vanish at $z = H$, corresponding to a stress free upper surface. Furthermore, there is no cell flux across the walls. Therefore, the boundary conditions for the governing system are given as follows,
\begin{equation}\label{13}
	\psi = 0 \quad \text{and} \quad \boldsymbol{J} \bcdot \hat{\boldsymbol{z}} = 0 \quad \text{at} \quad z = 0,H.
\end{equation}		 	
\begin{equation}\label{14}
	\frac{\partial \psi}{\partial z} =  0 \quad \text{at} \quad z = 0, \quad \text{and} \quad \zeta = 0 \quad \text{at} \quad z = H.
\end{equation}
The top boundary is exposed to a uniform oblique irradiation of magnitude \(I_t\). Both the top and bottom boundaries are assumed to be non-reflecting; therefore, the boundary conditions for the light intensity are given by 
\begin{equation}\label{15}
	\left.
	\begin{aligned}
		I(x, z, \theta, \phi) &= I_t \delta(\cos\theta + \cos\theta_0)\, \delta(\phi - \phi_0), && \pi/2 \leq \theta \leq \pi,\ 0 \leq \phi \leq 2\pi,\ \text{at } z = H, \\
		I(x, z, \theta, \phi) &= 0, && 0 \leq \theta \leq \pi/2,\ 0 \leq \phi \leq 2\pi,\ \text{at } z = 0
	\end{aligned}
	\right\}
\end{equation}

%%%%%%%%%%%%%%%%%%%%%%%%%%%%%%%%%%%%%%%%%%%%%%%%%%%%%%%%%%%%%%%%%%%%%%%%%%
\subsection{Scaling of the Equations}
To non-dimensionalize the system, all lengths are scaled on \(H\), the depth of the suspension layer, time is scaled by the diffusive time scale \(H^2/D\), the bulk fluid velocity on \(D/H\) and the cell concentration by \(\bar{n}\), the average cell concentration in the suspension~\cite{6ghorai2005}. For convenience, the non-dimensional variables are represented by the same symbols. Thus, the governing bioconvection equations in their dimensionless form are 
\begin{equation}\label{16}
	\boldsymbol{u} = (u,0,w) = \left(\frac{\partial \psi}{\partial z}, 0, \frac{-\partial \psi}{\partial x}\right), \quad \zeta = -\boldsymbol{\nabla}^2 \psi
\end{equation}	
\begin{equation}\label{17}
	Sc^{-1} \left( \frac{\partial \zeta}{\partial t}+\boldsymbol{\nabla}\cdot(\zeta\boldsymbol{u}) \right)= \boldsymbol{\nabla}^2\zeta  - R \, \frac{\partial n}{\partial x}.
\end{equation}	
\begin{equation}\label{18}
	\frac{\partial n}{\partial t} + \boldsymbol{\nabla} \bcdot \boldsymbol{J} = 0,
\end{equation}
where
\begin{equation*}
	\boldsymbol{J} = n \boldsymbol{u} +  n V_c \langle \boldsymbol{p} \rangle -  \boldsymbol{\nabla} n.
\end{equation*}			
The non-dimensional parameters considered in this study include the Schmidt number, \( S_c = \nu / D \) and the dimensionless swimming speed is given by \( V_c = W_c H / D \). The bioconvective Rayleigh number is defined as \( R = \bar{n} \vartheta g \Delta \rho H^{3} / \rho \nu D \).
	
In terms of the non-dimensional variables, the RTE becomes
\begin{equation}\label{19}
 \boldsymbol{s} \bcdot \boldsymbol{\nabla} I(\boldsymbol{x},\boldsymbol{s}) + \tau_H n I(\boldsymbol{x},\boldsymbol{s}) = \frac{\omega \tau_H n}{4\pi}\int_{0}^{4\pi}I(\boldsymbol{x},\boldsymbol{s}') (1 + A \cos\theta \cos\theta') d\Omega'
\end{equation}
where \(\tau_H = (\kappa + \varsigma) \bar{n} H \) is the dimensionless extinction coefficient. In terms of the direction cosines \((\xi,\nu) = (\sin\theta \cos\phi, \cos\theta)\) of the unit vector \(\boldsymbol{s}\), the dimensionless RTE can be written in the form 
\begin{equation}\label{20}
	\xi \frac{\partial I}{\partial x} + \nu \frac{\partial I}{\partial z} + \tau_H n I(\boldsymbol{x},\boldsymbol{s}) = \frac{\omega \tau_H n}{4\pi}\int_{0}^{4\pi}I(\boldsymbol{x},\boldsymbol{s}') (1 + A \cos\theta \cos\theta') d\Omega'
\end{equation}	
After scaling, the boundary conditions are 	
\begin{equation}\label{21}
	\psi = 0 \quad \text{and} \quad \boldsymbol{J} \bcdot \hat{\boldsymbol{z}} = 0 \quad \text{at} \quad z = 0,1.
\end{equation}		 	
\begin{equation}\label{22}
	\frac{\partial \psi}{\partial z} =  0 \quad \text{at} \quad z = 0, \quad \text{and} \quad \zeta = 0 \quad \text{at} \quad z = 1.
\end{equation}
Due to the scaling of the radiation intensity by the magnitude of the light intensity \(I_t\) at the top of the suspension, the corresponding non-dimensional boundary conditions for the intensity are given by 
\begin{equation}\label{23}
	\left.
	\begin{aligned}
		I(x, z, \theta, \phi) &= \delta(\cos\theta + \cos\theta_0)\, \delta(\phi - \phi_0), && \pi/2 \leq \theta \leq \pi,\ 0 \leq \phi \leq 2\pi,\ \text{at } z = 1, \\
		I(x, z, \theta, \phi) &= 0, && 0 \leq \theta \leq \pi/2,\ 0 \leq \phi \leq 2\pi,\ \text{at } z = 0
	\end{aligned}
	\right\}
\end{equation}
The phototaxis function \(T(G)\) is modeled as
\begin{equation}\label{24}
	T(G) = 0.8 \sin \left[\frac{3\pi}{2} \Xi(G) \right] - 0.1 \sin \left[\frac{\pi}{2} \Xi(G) \right],
\end{equation}
where \( \Xi(G) = 0.4\,G\,\exp[0.317\,(2.5-G)] \) with critical light intensity \(G_c = 1\).

%%%%%%%%%%%%%%%%%%%%%%%%%%%%%%%%%%%%%%%%%%%%%%%%%%%%%%%%%%%%%%%%%%%%%%%%%%%	 
	
\section{STABILITY ANALYSIS}
	
\subsection{The Basic State}
The steady-state solutions of Eqs.~(\ref{16})–(\ref{18}) and Eq.~(\ref{20}) subject to the boundary conditions must satisfy the following conditions
\begin{equation}\label{25}
\psi = 0, \quad \zeta = 0, \quad \boldsymbol{u} = 0, \quad n = n_b(z) \quad \text{and} \quad I = I_b(z,\theta)
\end{equation}
The total intensity \(G_b(z)\) and the radiative heat flux \(\boldsymbol{q_b}(z)\) in the steady state are given by
\begin{equation}\label{26}
	G_b(z) = \int_{0}^{4\pi} I_b(z,\theta) d \Omega \quad \text{and} \quad \boldsymbol{q_b}(z) = \int_{0}^{4\pi} I_b(z,\theta) \boldsymbol{s} d \Omega.
\end{equation} 
Since \(I_b(z,\theta)\) is independent of \(\phi\), so that the horizontal component of \(\boldsymbol{q_b}\) is zero. Therefore, \(\boldsymbol{q_b} = - q_b \hat{\boldsymbol{z}}\), where \(q_b = |\boldsymbol{q_b}|\).

The equation govern light intensity \(I_b(z,\theta)\) in the steady state is  
\begin{equation}\label{27}
	\nu \frac{\partial I_b}{\partial z} + \tau_H n_b(z) I_b(z,\theta) = \frac{\omega \tau_H n_b(z)}{4\pi}[G_b(z)-A\nu q_b(z)]
\end{equation}
We decompose the light intensity \(I_b(z,\theta)\) into two parts: the collimated component of the irradiation denoted by \(I_b^c(z,\theta)\) and the diffused part \(I_b^d(z,\theta)\), i.e., \(I_b(z,\theta) = I_b^c(z,\theta)+I_b^d(z,\theta)\). Then, the collimated part satisfies 
\begin{align}\label{28}
 	\frac{\partial I_b^{c}(z,\theta)}{\partial z} + \frac{\tau_H n_b(z)}{\nu} I_b^{c}(z,\theta) = 0
\end{align}
subject to the top boundary condition 
\begin{equation}\label{29}
 	I_b^{c}(1,\theta) =  \delta\,(\cos\theta+\cos\theta_0) \delta\, (\phi-\phi_0)
\end{equation}
On solving Eqs.~(\ref{28}) and (\ref{29}), we obtain
\begin{equation}\label{30}
	I_b^{c} = \exp\left(\int_{z}^{1}\frac{\tau_H n_b(z')}{\cos\theta} dz'\right) \delta\,(\cos\theta+\cos\theta_0) \delta\, (\phi-\phi_0)
\end{equation}
The governing equation for diffused part \(I_b^{d}(z,\theta)\) of the basic state light intensity  satisfies 
\begin{equation}\label{31}
	  \frac{\partial I_b^d}{\partial z} + \frac{\tau_H n_b(z)}{\nu} I_b^d(z,\theta) = \frac{\omega \tau_H n_b(z)}{4\pi \nu} [G_b(z) - A\nu q_b(z)]
\end{equation}
with boundary conditions 
\begin{equation}\label{32}
	\left.
	\begin{aligned}
		I_b^d(1,\theta) &= 0, && \pi/2 \leq \theta \leq \pi,\  \\
		I_b^d(0,\theta) &= 0, && 0 \leq \theta \leq \pi/2,\ 
	\end{aligned}
	\right\}
\end{equation}
The total intensity \(G_b(z)\) and radiative heat flux \(\boldsymbol{q_b}(z)\) is expressed as the sum of the collimated and diffused parts
\begin{equation}\label{33}
	G_b(z) = G_b^c(z) + G_b^d(z) \quad \text{and} \quad \boldsymbol{q_b}(z) = \boldsymbol{q_b}^c(z) + \boldsymbol{q_b}^d(z)
\end{equation}  
From Eq.~(\ref{30}), we get
\begin{equation}\label{34}
	G_b^{c} = \int_{0}^{4\pi} I_b^{c}(z,\theta) d\Omega = \exp\left(-\int_{z}^{1}\frac{\tau_H n_b(z')}{\cos\theta_0} dz'\right)
\end{equation}
and
\begin{equation}\label{35}
	\boldsymbol{q_b}^{c} = \int_{0}^{4\pi} I_b^{c}(z,\theta) \boldsymbol{s} d\Omega = -\cos \theta_0 \exp\left(-\int_{z}^{1}\frac{\tau_H n_b(z')}{\cos\theta_0} dz'\right)\hat{\boldsymbol{z}}
\end{equation}
The diffused part of basic total light intensity \(G_b(z)\) and radiative heat flux \(\boldsymbol{q_b}(z)\) satisfies 
\begin{equation}\label{36}
	G_b^{d} = \int_{0}^{4\pi} I_b^{d}(z,\theta) d\Omega, \quad \text{and} \quad \boldsymbol{q_b}^{d} = \int_{0}^{4\pi} I_b^{d}(z,\theta) \boldsymbol{s} d\Omega
\end{equation}
Now, we define a new variable 
\begin{equation}\label{37}
	\tau = \int_{z}^{1}\tau_H n_b(z') dz'
\end{equation}	
In terms of the variable \(\tau\), Eq.~(\ref{31}) for the outgoing \((0 \leq \theta \leq \pi/2)\) light intensity, \(I_b^{d^+}(\tau,\nu)\), is written as
\begin{equation}\label{38}
	\frac{\partial I_b^{d^+}}{\partial\tau} - \frac{1}{\nu} I_b^{d^+}(\tau
	,\nu) = -\frac{\omega}{4\pi \nu}[G_b(\tau)-A \,q_b(\tau) \nu], \quad 0 \leq \nu \leq 1	
\end{equation}
with boundary condition
\begin{equation}\label{39}
	I_b^{d^+}(\tau_H,\nu) =  0, \quad 0 \leq \nu \leq 1.
\end{equation}
Solving Eqs.~(\ref{38}) and (\ref{39}), we get
\begin{equation}\label{40}
	 I_b^{d^+}(\tau,\nu) = -\frac{\omega}{4\pi\nu}\int_{\tau_H}^{\tau}e^{(\tau-\tau')/\nu}[G_b(\tau')-A q_b(\tau') \nu]  d\tau', \quad 0 \leq \nu \leq 1	
\end{equation}
Similarly, the incoming \((\pi/2 \leq \theta \leq \pi)\) light intensity, \(I_b^{d^-}(\tau,\nu)\) is given by
\begin{equation}\label{41}
	I_b^{d^-}(\tau,\nu) = -\frac{\omega}{4\pi\nu}\int_{0}^{\tau}e^{(\tau-\tau')/\nu}[G_b(\tau')-A\, q_b(\tau') \nu]  d\tau', \quad -1 \leq \nu \leq 0	
\end{equation}
Substituting the Eqs.~(\ref{40}) and (\ref{41}) into Eq.~(\ref{33}), we get two coupled Fredholm integral equations 
\begin{align}\label{42}
	G_b(\tau) = e^{(-\tau/\cos\theta_0)} +  \frac{\omega}{2}\int_{0}^{\tau_H}[G_b(\tau')E_1(|\tau-\tau'|)+A\,sgn(\tau-\tau')q_b(\tau')E_2(|\tau-\tau'|)]d\tau'
\end{align}
\begin{align}\label{43}
	q_b(\tau) = \cos\theta_0 e^{(-\tau/\cos\theta_0)} + \frac{\omega}{2}\int_{0}^{\tau_H}[A\,q_b(\tau')E_3(|\tau-\tau'|)+sgn(\tau-\tau')G_b(\tau')E_2(|\tau-\tau'|)]d\tau'
\end{align}
where, `\(sgn(x)\)' is the sign function and \(E_n(x)\) is the exponential integral function of order \(n\).
Eqs.~(\ref{42}) and (\ref{43}) are solved using the subtraction of singularity method.
From Eq.~(\ref{11}), the average swimming direction in the steady state becomes
\begin{equation}\label{44}
	\langle \boldsymbol{p_b} \rangle = -T(G_b) \frac{\boldsymbol{q}_b}{q_b} = T(G_b)\hat{\boldsymbol{z}}.
\end{equation}
so that the steady-state concentration \( n_b(z) \) satisfies the equation	
\begin{equation}\label{45}
	\frac{\ud n_b}{\ud z} - V_c T_b n_b = 0,
\end{equation}	
where \( T_b = T(G_b) \) represents the phototaxis function evaluated at steady-state light intensity \( G_b \).	    
Eq.~(\ref{45}) is supplemented by the cell conservation relation 
\begin{equation}\label{46}
	\int_0^1 n_b(z)(\boldsymbol{z}) = 1
\end{equation}	
Next, Eqs.~(\ref{42})-(\ref{46}) constitute a boundary value problem which is numerically solved by using a shooting method.

%%%%%%%%%%%%%%%%%%%%%%%%%%%%%%%%%%%%%%%%%%%%%%%%%%%%%%%%%%%%%%%%%%%%%%%	   
	
\subsection{Linear Stability Analysis}
To analyze the stability of the system, a small perturbation of amplitude \( \epsilon \) (where \( 0 < \epsilon \ll 1 \)) is introduced to the steady-state solution, leading to	
\begin{equation}\label{47}
	\begin{pmatrix}
		\psi \\ \zeta \\ \boldsymbol{u} \\ n \\ I
	\end{pmatrix}
		=
	\begin{pmatrix}
		0 + \epsilon \psi_1 + O(\epsilon^{2}) \\ 0 + \epsilon \zeta_1 + O(\epsilon^{2}) \\ 0 + \epsilon \boldsymbol{u}_1 + O(\epsilon^{2}) \\ n_b + \epsilon n_1 + O(\epsilon^{2}) \\ I_b + \epsilon I_1 + O(\epsilon^{2}) 
		\end{pmatrix},
\end{equation}	 
	
Substituting these perturbed quantities into Eqs.~(\ref{16})–(\ref{18}) and retaining only terms of order \( O(\epsilon) \) results in the following linearized system		
\begin{equation}\label{48}
	Sc^{-1} \frac{\partial (\boldsymbol{\nabla}^2 \psi_1)}{\partial t} = \boldsymbol{\nabla}^4 \psi_1 + R\,\frac{\partial n_1}{\partial x},
\end{equation}	
\begin{equation}\label{49}
	\frac{\partial n_1}{\partial t} + V_c \boldsymbol{\nabla} \cdot \left( \langle \boldsymbol{p_b} \rangle n_1 + \langle \boldsymbol{p_1} \rangle n_b \right) - \boldsymbol{\nabla}^2 n_1 = \frac{\partial \psi_1}{\partial x} \frac{\ud n_b}{\ud z}.
\end{equation}	
Let \(G_1\) be the total perturbed radiation intensity, i.e., \(G=G_b+\epsilon G_1 + O(\epsilon^2)\). Since \(G_b=G_b^{c}+G_b^{d}\), so that 
\begin{equation}\label{50}
	G=(G_b^{c}+\epsilon G_1^{c}) + (G_b^{d}+\epsilon G_1^{d}) + O(\epsilon^2)
\end{equation}
then, we have 
\begin{equation}\label{51}
	G_1 = G_1^{c} + G_1^{d}
\end{equation}
where
\begin{equation*}
	G_1^{c} = \left(\int_{1}^{z}\frac{\tau_H n_1}{\cos\theta_0} dz'\right) \exp\left(-\int_{z}^{1}\frac{\tau_H n_b(z')}{\cos\theta_0} dz'\right) \quad \text{and} \quad G_1^{d} = \int_{0}^{4\pi} I_1^{d}(\boldsymbol{x},\boldsymbol{s}) d\Omega
\end{equation*}	
Similarly, the radiative heat flux \(\boldsymbol{q}=\boldsymbol{q_b}+\epsilon \boldsymbol{q_1} + O(\epsilon^2)\), then the perturbed total radiative heat flux
\begin{equation}\label{52}
	\boldsymbol{q_1} = \boldsymbol{q_1}^{c} + \boldsymbol{q_1}^{d}
\end{equation}
where
\begin{equation*}
	\boldsymbol{q_1}^{c} = -\cos\theta_0 \left(\int_{1}^{z}\frac{\tau_H n_1}{\cos\theta_0} dz'\right) \exp\left(-\int_{z}^{1}\frac{\tau_H n_b(z')}{\cos\theta_0} dz'\right) \hat{\boldsymbol{z}} \quad \text{and} \quad \boldsymbol{q_1}^{d} = \int_{0}^{4\pi} I_1^{d}(\boldsymbol{x},\boldsymbol{s}) \boldsymbol{s} d\Omega
\end{equation*} 
After solving the expression 
\begin{equation*}
	-T(G_b+\epsilon G_1 + O(\epsilon^2)) \frac{\boldsymbol{q_b}+\epsilon \boldsymbol{q_1} + O(\epsilon^2)}{|\boldsymbol{q_b}+\epsilon \boldsymbol{q_1} + O(\epsilon^2)|} - T(G_b)\hat{\boldsymbol{z}},
\end{equation*}
the perturbed swimming orientation is 
\begin{equation}\label{53}
	\langle \boldsymbol{p_1} \rangle = G_1 \frac{\ud T_b}{\ud G} \hat{\boldsymbol{z}} - T(G_b) \frac{\boldsymbol{q_1}^{H}}{q_b}
\end{equation}
where \(\boldsymbol{q_1}^{H}\) is the horizontal component of the perturbed radiative heat flux \(\boldsymbol{q_1}\). 
Now, the Eqs.~(\ref{48}) and (\ref{49}) are simplified into a pair of equations involving \(\psi_1\) and \(n_1\) and these perturbed quantities can be decomposed into normal modes such that 
\begin{equation}\label{54}
	\psi_1 = \Psi(z)\,e^{\gamma \,t + ikx} \quad \text{and} \quad n_1 = \Theta(z)\, e^{\gamma \,t + ikx}
\end{equation}
where \(\gamma\) is the growth rate and \(k\) is the non-dimensional horizontal wave number.
The diffuse component of the perturbed intensity \(I_1^{d}\) satisfies
\begin{equation}\label{55}
	\xi \frac{\partial I_1^{d}}{\partial x} + \nu \frac{\partial I_1^{d}}{\partial z} + \tau_H n_b I_1^{d}(\boldsymbol{x},\boldsymbol{s}) = \frac{\omega \tau_H }{4\pi} \left( n_b G_1 + n_1 G_b + A \nu (n_b \boldsymbol{q_1} \cdot \hat{\boldsymbol{z}} - q_b n_1) \right) - \tau_H n_1 I_b
\end{equation}
subject to the boundary conditions 
\begin{equation}\label{56}
	\left.
	\begin{aligned}
		I_1^d(x, 1, \theta, \phi) &= 0, && \pi/2 \leq \theta \leq \pi,\ 0 \leq \phi \leq 2\pi \\
		I_1^d(x, 0, \theta, \phi) &= 0, && 0 \leq \theta \leq \pi/2,\ 0 \leq \phi \leq 2\pi
	\end{aligned}
	\right\}
\end{equation}
The perturbed intensity \(I_1^{d}\) can be represented by the following expression
\begin{equation}\label{57}
	I_1^{d} = \Phi^{d}(z,\xi,\nu) e^{(\gamma \,t + ikx)}
\end{equation}
From Eq.(\ref{51}), we get
\begin{equation}\label{58}
	G_1^{c} = \left(\frac{\tau_H}{\cos\theta_0}\int_{1}^{z}\Theta(z') dz'\right) \exp\left(-\frac{\tau_H}{\cos\theta_0}\int_{z}^{1} n_b(z') dz'\right) e^{(\gamma \,t + ikx)} = g_1^{c}(z) e^{(\gamma \,t + ikx)}
\end{equation}
and 
\begin{equation}\label{59}
	G_1^{d} = g_1^{d}(z) e^{(\gamma \,t + ikx)} = \left( \int_{0}^{4 \pi} \Phi^{d}(z,\xi,\nu) d \Omega \right) e^{(\gamma \,t + ikx)}
\end{equation}
Here, \(g_1(z)\) is the perturbed total intensity, expressed as the sum of \(g_1^{c}(z)\) and \(g_1^{d}(z)\). Similarly, from Eq.~(\ref{52}), we have
\begin{equation}\label{60}
	\boldsymbol{q_1} = (q_1^{x},q_1^{z}) = [P_x(z),P_z(z)] e^{(\gamma \,t + ikx)}
\end{equation}
where
\begin{equation}\label{61}
	P_x(z) = \int_{0}^{4 \pi} \Phi^{d}(z,\xi,\nu) \xi d \Omega \quad \text{and} \quad P_z(z) = -g_1^c(z) + \int_{0}^{4 \pi} \Phi^{d}(z,\xi,\nu) \nu d \Omega
\end{equation}
From Eq.~\ref{55} , \(\Phi^{d}\) satisfies 
\begin{equation}\label{62}
	\frac{d \Phi^{d}}{dz} + \frac{[ik \xi +  \tau_H n_b]}{\nu} \Phi^{d} = \frac{\omega \tau_H}{4 \pi} [n_b g_1 + G_b \Theta + A\, \nu (n_b P_z(z) - q_b \Theta)] - \frac{\tau_H}{\nu} I_b \Theta(z)
\end{equation}
with boundary conditions 
\begin{equation}\label{63}
	\left.
	\begin{aligned}
		\Phi^d(1, \xi, \nu) &= 0, && \pi/2 \leq \theta \leq \pi,\ 0 \leq \phi \leq 2\pi \\
		\Phi^d(0, \xi, \nu) &= 0, && 0 \leq \theta \leq \pi/2,\ 0 \leq \phi \leq 2\pi
	\end{aligned}
	\right\}
\end{equation}
Thus, the governing equations become	
\begin{equation}\label{64}
	 \left( \gamma \, S_c^{-1} + k^2 - \frac{\ud^2}{\ud z^2} \right) \left( \frac{\ud^2}{\ud z^2} - k^2 \right) \Psi(z) = R (ik) \Theta(z),	
\end{equation}
\begin{equation}\label{65}
	\left( \gamma  + k^2 - \frac{\ud^2}{\ud z^2} \right) \Theta + V_c \frac{\ud}{\ud z} \left( T_b\,\Theta + n_b \frac{\ud T_b}{\ud G} g_1 \right) - \frac{i k V_c n_b T_b P_x(z)}{q_b} = (ik) \frac{\ud n_b}{\ud z} \Psi(z)	
\end{equation}	
subject to the boundary conditions 	
\begin{equation}\label{66}
	\Psi = \frac{\ud \Psi}{\ud z}  = \frac{d\Theta}{dz} - V_c T_b \Theta - V_c \,n_b \frac{\ud T_b}{\ud G}\,g_1 = 0 \quad \text{on} \quad z = 0,
\end{equation}	
\begin{equation}\label{67}
	\Psi = \frac{\ud^2 \Psi}{\ud z^2} = \frac{d\Theta}{dz} - V_c T_b \Theta - V_c \,n_b \frac{\ud T_b}{\ud G}\,g_1 = 0 \quad \text{on} \quad z = 1.
\end{equation}
The system forms an eigenvalue problem for \( \gamma \), determining the stability of the perturbations.	Eq.~(\ref{65}) represents a linear integro-differential equation with non-constant coefficients arising from the functional form of \(g_1\) so that we introduce a new variable	
\begin{equation}\label{68}
	\varphi(z) = \int_1^z \Theta(z') dz'.
\end{equation}	
The perturbed equations, rewritten using \( D \equiv d/dz \), take the following form	
\begin{equation}\label{69}
	 ( \gamma \, S_c^{-1} + k^{2} - D^{2})( D^{2} - k^{2}) \Psi(z) = R (ik) D\varphi,	
\end{equation}
\begin{equation}\label{70}
	D^3\varphi - V_c T_b D^2\varphi - [\gamma + k^2 + \Gamma_3(z)] D\varphi - \Gamma_2(z) \varphi - \Gamma_1(z) = -(ik) Dn_b \Psi(z)
\end{equation}
where
\begin{equation*}
	\Gamma_1 = V_c \frac{\ud}{\ud z} \left(n_b \frac{\ud T_b}{\ud G} g_1^{d} \right) - \frac{i k V_c n_b T_b P_x(z)}{q_b},
\end{equation*}	
\begin{equation*}
	\Gamma_2 = (\tau_H/\cos\theta_0) V_c \frac{\ud}{\ud z} \left(n_b G_b^{c} \frac{\ud T_b}{\ud G} \right),
\end{equation*}	
\begin{equation*}
	\Gamma_3 = 2(\tau_H/\cos\theta_0) V_c n_b G_b^{c} \frac{\ud T_b}{\ud G} + V_c \frac{\ud T_b}{\ud G} \frac{\ud G_b^{d}}{\ud z}.
\end{equation*}	
The boundary conditions become
\begin{equation}\label{71}
	\Psi = D\Psi  = D^2\varphi - V_c T_b D\varphi - V_c \,n_b \frac{\ud T_b}{\ud G}\,g_1 = 0 \quad \text{on} \quad z = 0,
\end{equation}	
\begin{equation}\label{72}
	\Psi = D^2\Psi = D^2\varphi - V_c T_b D\varphi - V_c \,n_b \frac{\ud T_b}{\ud G}\,g_1 = 0 \quad \text{on} \quad z = 1.
\end{equation}
and there is an extra boundary condition	
\begin{equation}\label{73}
	\varphi(z) = 0 \quad \text{on} \quad z = 1.
\end{equation}
which directly follows from Eq.~(\ref{68}). \\
The system governed by Eqs.~(\ref{69})–(\ref{73}) forms an eigen-value problem for \(\gamma\), determining the stability of the disturbances. To analyze the linear stability of this system, we used a fourth-order accurate finite difference scheme based on the Newton-Raphson-Kantorovich (NRK) iterations~\cite{11cash1980}. This method facilitates the computation of the growth rate, Re$(\gamma)$, and calculation of the neutral stability curves in the $(k, R)$-plane for specific parameter settings. A neutral curve is defined as the locus of points where \( \text{Re}(\gamma) = 0 \). If in addition \( \text{Im}(\gamma) = 0 \) along this curve, then the principle of exchange of stabilities is said to be valid and the disturbance is stationary or non-oscillatory. Alternatively, if \( \text{Im}(\gamma) \neq 0 \), then the oscillatory or overstable solutions exist for governing system. For a fixed set of parameters, the neutral curve, denoted by \( R^{(n)}(k) \), where \( n = 1, 2, 3, \dots \), has an infinite number of branches, each of which represents a distinct solution to the linear stability problem. The solution branch of most interest is that on which the critical value \(R_c\) of \(R\) occurs at the critical wave number \(k_c\). The most unstable solution corresponds to this branch and the wavelength of initial perturbation can be calculated by using the expression \( \lambda_c = \frac{2\pi}{k_c} \).
	
%%%%%%%%%%%%%%%%%%%%%%%%%%%%%%%%%%%%%%%%%%%%%%%%%%%%%%%%%%%%%%%%%%%%%	
\section{NUMERICAL RESULTS}
	
%This study systematically explores the parameter space to enhance the understanding the effects of water absorption on algal suspensions by maintaining specific parameters constant while varying others. This approach enables a focused investigation of particular system dynamics, revealing the distinct influences on bioconvection onset. The governing parameters are fixed at \( Sc = 20 \), \( \mathrm{I_s} = 0.8 \) and \(\alpha_0 = 0.006 \) to ensure consistency, while other parameters are varied to examine their effects. The values \( \alpha_1 = 0.5 \) and \( \alpha_1 = 1.0 \) are considered to account for differences in the light absorption properties of microorganisms. Additionally, \( G_c \) is selected such that the steady state occurs at the top, three-quarters, or mid-height of the suspension. The primary objective is to determine the critical bioconvective Rayleigh number, \( R_c \), corresponding to critical wave number, \( k_c \).

To investigate the dynamic behavior of the system and its influence on the onset of bioconvection, specific parameters are held constant while others are systematically varied. In addition, the parameter \( G_c \) is chosen such that the steady-state cell concentration occurs near the top, at three-quarters height, or at mid-depth of the suspension. The fixed parameters for the study are given as follows: \(G_c =1.0, S_c = 20\) while the varying parameters are \(\tau_H = 0.5,1.0\), the optical thickness, \(0 \leq \omega \leq 1\), single scattering albedo, \(V_c = 10,15,20\), scaled swimming speed, and \(A = 0,0.4,0.8\), forward scattering coefficient. The position of the steady state concentration peak varies with the angle of incidence,\(\theta_i\), occurring at the mid-height, near the three-quarter height, and near the top of the suspension for \(\theta_i = 0, 40, 80\), respectively.
	
\begin{figure}[!htbp]
	\centering
	\includegraphics[scale=0.63]{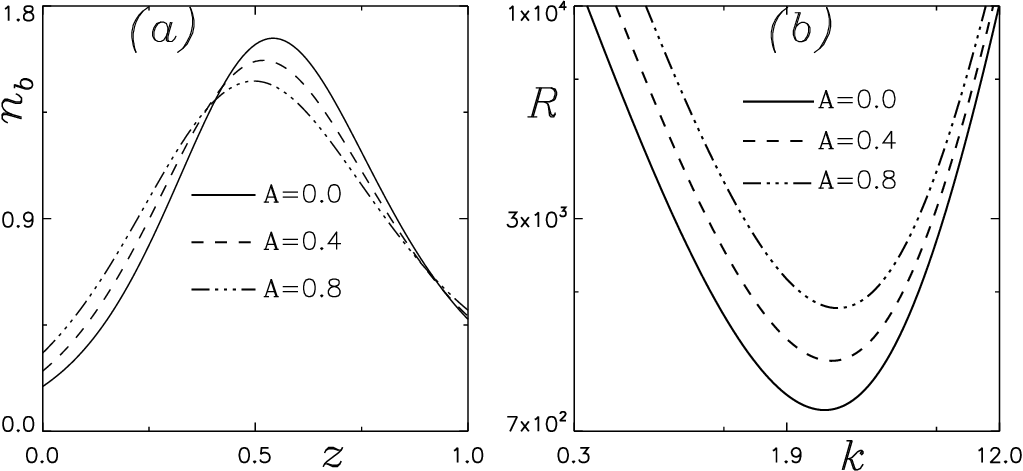}
	\caption{(a) Concentration profiles at the steady state, and (b) the corresponding neutral curves for varying forward scattering coefficient \(A\), where fixed parameters are \(\theta_i = 0, V_c = 15, \tau_H = 0.5, G_c = 1.0, \omega=0.469\).}
	\label{fig2}
\end{figure}

Fig.(\ref{fig2}) illustrates the effect of linearly anisotropic scattering coefficient \(A\) on the basic equilibrium concentration profile along with  the corresponding modes of instability. When forward scattering coefficient \(A=0\), the maximum basic concentration occurs near the mid-height of the suspension and the concentration peak shifts in the downward direction as \(A\) increases. As a result, the critical Rayleigh number \(R_c\) increases which indicates that the suspension becomes more stable for higher forward scattering coefficients.

%%%%%%%%%%%%%%%%%%%%%%%%%%%%%%%%%%%%%%%%%%%%%%%%%%%%%%%%%%%%%
\section{Conclusion}
	
%This study presents a novel thermal phototactic bioconvection model that integrates thermal effects into the dynamics of bioconvective systems in a porous medium. The suspension is exposed to collimated light from above and subjected to heating or cooling from below. The primary objective is to examine how thermal variations impact the onset and behavior of phototactic bioconvection in a porous environment. The numerical results, obtained through a linear stability analysis under specific control parameters, are summarized below.
	
The accumulation of algal cells within the suspension forms a sublayer, whose position is governed by the critical light intensity. When the critical intensity matches the incident light intensity, cells cluster at the top; as the critical intensity decreases, this aggregation shifts downward. Moreover, a reduction in light intensity corresponds to a decrease in the maximum concentration at the sublayer. 
	
The linear stability analysis reveals both steady and oscillatory solutions, with oscillations becoming particularly pronounced when the microorganism layer is positioned around three-quarters of the suspension height, dictated by the critical light intensity. 
%A distinct relationship emerges: variations in the critical Rayleigh number for bioconvection correspond to similar changes in the critical thermal Rayleigh number. This indicates that heating from below destabilizes the suspension, whereas cooling has a stabilizing effect. Additionally, when microorganisms accumulate near the top boundary, the critical pattern wavelength increases as the critical thermal Rayleigh number rises, while a decrease in the critical light intensity leads to a reduction in the critical pattern wavelength under increasing thermal Rayleigh numbers.

%%%%%%%%%%%%%%%%%%%%%%%%%%%%-OM NAMAH SHIVAY-%%%%%%%%%%%	
%\section*{AUTHOR DECLARATIONS}
\section*{CONFLICT OF INTEREST}
The authors state that they have no conflicts of interest.
\section*{DATA AVAILABILITY}
The findings of this study are supported by the data contained within this article.
\nocite{*}
\section*{REFERENCES}
\bibliography{ANISO_OBLIQUE}

\providecommand{\noopsort}[1]{}\providecommand{\singleletter}[1]{#1}%
\begin{thebibliography}{25}%
\makeatletter
\providecommand \@ifxundefined [1]{%
 \@ifx{#1\undefined}
}%
\providecommand \@ifnum [1]{%
 \ifnum #1\expandafter \@firstoftwo
 \else \expandafter \@secondoftwo
 \fi
}%
\providecommand \@ifx [1]{%
 \ifx #1\expandafter \@firstoftwo
 \else \expandafter \@secondoftwo
 \fi
}%
\providecommand \natexlab [1]{#1}%
\providecommand \enquote  [1]{``#1''}%
\providecommand \bibnamefont  [1]{#1}%
\providecommand \bibfnamefont [1]{#1}%
\providecommand \citenamefont [1]{#1}%
\providecommand \href@noop [0]{\@secondoftwo}%
\providecommand \href [0]{\begingroup \@sanitize@url \@href}%
\providecommand \@href[1]{\@@startlink{#1}\@@href}%
\providecommand \@@href[1]{\endgroup#1\@@endlink}%
\providecommand \@sanitize@url [0]{\catcode `\\12\catcode `\$12\catcode
  `\&12\catcode `\#12\catcode `\^12\catcode `\_12\catcode `\%12\relax}%
\providecommand \@@startlink[1]{}%
\providecommand \@@endlink[0]{}%
\providecommand \url  [0]{\begingroup\@sanitize@url \@url }%
\providecommand \@url [1]{\endgroup\@href {#1}{\urlprefix }}%
\providecommand \urlprefix  [0]{URL }%
\providecommand \Eprint [0]{\href }%
\providecommand \doibase [0]{http://dx.doi.org/}%
\providecommand \selectlanguage [0]{\@gobble}%
\providecommand \bibinfo  [0]{\@secondoftwo}%
\providecommand \bibfield  [0]{\@secondoftwo}%
\providecommand \translation [1]{[#1]}%
\providecommand \BibitemOpen [0]{}%
\providecommand \bibitemStop [0]{}%
\providecommand \bibitemNoStop [0]{.\EOS\space}%
\providecommand \EOS [0]{\spacefactor3000\relax}%
\providecommand \BibitemShut  [1]{\csname bibitem#1\endcsname}%
\let\auto@bib@innerbib\@empty
%</preamble>
\bibitem [{\citenamefont {Platt}(1961)}]{12platt1961}%
  \BibitemOpen
  \bibfield  {author} {\bibinfo {author} {\bibfnamefont {J.~R.}\ \bibnamefont
  {Platt}},\ }\href@noop {} {\bibfield  {journal} {\bibinfo  {journal}
  {Science}\ }\textbf {\bibinfo {volume} {133}},\ \bibinfo {pages} {1766}
  (\bibinfo {year} {1961})}\BibitemShut {NoStop}%
\bibitem [{\citenamefont {Pedley}\ and\ \citenamefont
  {Kessler}(1992)}]{13pedley1992}%
  \BibitemOpen
  \bibfield  {author} {\bibinfo {author} {\bibfnamefont {T.~J.}\ \bibnamefont
  {Pedley}}\ and\ \bibinfo {author} {\bibfnamefont {J.~O.}\ \bibnamefont
  {Kessler}},\ }\href@noop {} {\bibfield  {journal} {\bibinfo  {journal}
  {Annual Review of Fluid Mechanics}\ }\textbf {\bibinfo {volume} {24}},\
  \bibinfo {pages} {313} (\bibinfo {year} {1992})}\BibitemShut {NoStop}%
\bibitem [{\citenamefont {Hill}\ and\ \citenamefont
  {Pedley}(2005)}]{14hill2005}%
  \BibitemOpen
  \bibfield  {author} {\bibinfo {author} {\bibfnamefont {N.~A.}\ \bibnamefont
  {Hill}}\ and\ \bibinfo {author} {\bibfnamefont {T.~J.}\ \bibnamefont
  {Pedley}},\ }\href@noop {} {\bibfield  {journal} {\bibinfo  {journal} {Fluid
  Dynamics Research}\ }\textbf {\bibinfo {volume} {37}},\ \bibinfo {pages} {1}
  (\bibinfo {year} {2005})}\BibitemShut {NoStop}%
\bibitem [{\citenamefont {Kessler}(1985)}]{1kessler1985}%
  \BibitemOpen
  \bibfield  {author} {\bibinfo {author} {\bibfnamefont {J.~O.}\ \bibnamefont
  {Kessler}},\ }\href@noop {} {\bibfield  {journal} {\bibinfo  {journal}
  {Contemporary Physics}\ }\textbf {\bibinfo {volume} {26}},\ \bibinfo {pages}
  {147} (\bibinfo {year} {1985})}\BibitemShut {NoStop}%
\bibitem [{\citenamefont {Wager}(1911)}]{21wager1911vii}%
  \BibitemOpen
  \bibfield  {author} {\bibinfo {author} {\bibfnamefont {H.~W.~T.}\
  \bibnamefont {Wager}},\ }\href@noop {} {\bibfield  {journal} {\bibinfo
  {journal} {Philosophical Transactions of the Royal Society of London. Series
  B, Containing Papers of a Biological Character}\ }\textbf {\bibinfo {volume}
  {201}},\ \bibinfo {pages} {333} (\bibinfo {year} {1911})}\BibitemShut
  {NoStop}%
\bibitem [{\citenamefont {Vincent}(1995)}]{22vincent1995mathematical}%
  \BibitemOpen
  \bibfield  {author} {\bibinfo {author} {\bibfnamefont {R.~V.}\ \bibnamefont
  {Vincent}},\ }\emph {\bibinfo {title} {Mathematical modelling of phototaxis
  in motile microorganisms}},\ \href@noop {} {Ph.D. thesis},\ \bibinfo
  {school} {University of Leeds (Department of Applied Mathematical Studies)}
  (\bibinfo {year} {1995})\BibitemShut {NoStop}%
\bibitem [{\citenamefont {H{\"a}der}(1987)}]{3hader1987}%
  \BibitemOpen
  \bibfield  {author} {\bibinfo {author} {\bibfnamefont {D.-P.}\ \bibnamefont
  {H{\"a}der}},\ }\href@noop {} {\bibfield  {journal} {\bibinfo  {journal}
  {Archives of microbiology}\ }\textbf {\bibinfo {volume} {147}},\ \bibinfo
  {pages} {179} (\bibinfo {year} {1987})}\BibitemShut {NoStop}%
\bibitem [{\citenamefont {Straughan}(1993)}]{5straughan1993}%
  \BibitemOpen
  \bibfield  {author} {\bibinfo {author} {\bibfnamefont {B.}~\bibnamefont
  {Straughan}},\ }\href@noop {} {\emph {\bibinfo {title} {Mathematical aspects
  of penetrative convection}}}\ (\bibinfo  {publisher} {CRC Press},\ \bibinfo
  {year} {1993})\BibitemShut {NoStop}%
\bibitem [{\citenamefont {Williams}\ and\ \citenamefont
  {Bees}(2011)}]{2williams2011}%
  \BibitemOpen
  \bibfield  {author} {\bibinfo {author} {\bibfnamefont {C.~R.}\ \bibnamefont
  {Williams}}\ and\ \bibinfo {author} {\bibfnamefont {M.~A.}\ \bibnamefont
  {Bees}},\ }\href@noop {} {\bibfield  {journal} {\bibinfo  {journal} {Journal
  of Experimental Biology}\ }\textbf {\bibinfo {volume} {214}},\ \bibinfo
  {pages} {2398} (\bibinfo {year} {2011})}\BibitemShut {NoStop}%
\bibitem [{\citenamefont {Vincent}\ and\ \citenamefont
  {Hill}(1996)}]{8vincent1996}%
  \BibitemOpen
  \bibfield  {author} {\bibinfo {author} {\bibfnamefont {R.~V.}\ \bibnamefont
  {Vincent}}\ and\ \bibinfo {author} {\bibfnamefont {N.~A.}\ \bibnamefont
  {Hill}},\ }\href@noop {} {\bibfield  {journal} {\bibinfo  {journal} {Journal
  of Fluid Mechanics}\ }\textbf {\bibinfo {volume} {327}},\ \bibinfo {pages}
  {343} (\bibinfo {year} {1996})}\BibitemShut {NoStop}%
\bibitem [{\citenamefont {Ghorai}\ and\ \citenamefont
  {Hill}(2005)}]{6ghorai2005}%
  \BibitemOpen
  \bibfield  {author} {\bibinfo {author} {\bibfnamefont {S.}~\bibnamefont
  {Ghorai}}\ and\ \bibinfo {author} {\bibfnamefont {N.~A.}\ \bibnamefont
  {Hill}},\ }\href@noop {} {\bibfield  {journal} {\bibinfo  {journal} {Physics
  of fluids}\ }\textbf {\bibinfo {volume} {17}},\ \bibinfo {pages} {074101}
  (\bibinfo {year} {2005})}\BibitemShut {NoStop}%
\bibitem [{\citenamefont {Ghorai}\ \emph {et~al.}(2010)\citenamefont {Ghorai},
  \citenamefont {Panda},\ and\ \citenamefont {Hill}}]{4ghorai2010}%
  \BibitemOpen
  \bibfield  {author} {\bibinfo {author} {\bibfnamefont {S.}~\bibnamefont
  {Ghorai}}, \bibinfo {author} {\bibfnamefont {M.~K.}\ \bibnamefont {Panda}}, \
  and\ \bibinfo {author} {\bibfnamefont {N.~A.}\ \bibnamefont {Hill}},\
  }\href@noop {} {\bibfield  {journal} {\bibinfo  {journal} {Physics of
  Fluids}\ }\textbf {\bibinfo {volume} {22}},\ \bibinfo {pages} {071901}
  (\bibinfo {year} {2010})}\BibitemShut {NoStop}%
\bibitem [{\citenamefont {Ghorai}\ and\ \citenamefont
  {Panda}(2013)}]{23ghorai2013}%
  \BibitemOpen
  \bibfield  {author} {\bibinfo {author} {\bibfnamefont {S.}~\bibnamefont
  {Ghorai}}\ and\ \bibinfo {author} {\bibfnamefont {M.}~\bibnamefont {Panda}},\
  }\href@noop {} {\bibfield  {journal} {\bibinfo  {journal} {European Journal
  of Mechanics-B/Fluids}\ }\textbf {\bibinfo {volume} {41}},\ \bibinfo {pages}
  {81} (\bibinfo {year} {2013})}\BibitemShut {NoStop}%
\bibitem [{\citenamefont {Panda}\ and\ \citenamefont
  {Ghorai}(2013)}]{9panda2013}%
  \BibitemOpen
  \bibfield  {author} {\bibinfo {author} {\bibfnamefont {M.~K.}\ \bibnamefont
  {Panda}}\ and\ \bibinfo {author} {\bibfnamefont {S.}~\bibnamefont {Ghorai}},\
  }\href@noop {} {\bibfield  {journal} {\bibinfo  {journal} {Physics of
  Fluids}\ }\textbf {\bibinfo {volume} {25}},\ \bibinfo {pages} {071902}
  (\bibinfo {year} {2013})}\BibitemShut {NoStop}%
\bibitem [{\citenamefont {Panda}\ and\ \citenamefont
  {Singh}(2016)}]{7panda2016}%
  \BibitemOpen
  \bibfield  {author} {\bibinfo {author} {\bibfnamefont {M.~K.}\ \bibnamefont
  {Panda}}\ and\ \bibinfo {author} {\bibfnamefont {R.}~\bibnamefont {Singh}},\
  }\href@noop {} {\bibfield  {journal} {\bibinfo  {journal} {Physics of
  Fluids}\ }\textbf {\bibinfo {volume} {28}},\ \bibinfo {pages} {054105}
  (\bibinfo {year} {2016})}\BibitemShut {NoStop}%
\bibitem [{\citenamefont {Panda}\ \emph {et~al.}(2016)\citenamefont {Panda},
  \citenamefont {Singh}, \citenamefont {Mishra},\ and\ \citenamefont
  {Mohanty}}]{24panda2016}%
  \BibitemOpen
  \bibfield  {author} {\bibinfo {author} {\bibfnamefont {M.}~\bibnamefont
  {Panda}}, \bibinfo {author} {\bibfnamefont {R.}~\bibnamefont {Singh}},
  \bibinfo {author} {\bibfnamefont {A.~C.}\ \bibnamefont {Mishra}}, \ and\
  \bibinfo {author} {\bibfnamefont {S.~K.}\ \bibnamefont {Mohanty}},\
  }\href@noop {} {\bibfield  {journal} {\bibinfo  {journal} {Physics of
  Fluids}\ }\textbf {\bibinfo {volume} {28}} (\bibinfo {year}
  {2016})}\BibitemShut {NoStop}%
\bibitem [{\citenamefont {Panda}(2020)}]{25panda2020}%
  \BibitemOpen
  \bibfield  {author} {\bibinfo {author} {\bibfnamefont {M.~K.}\ \bibnamefont
  {Panda}},\ }\href@noop {} {\bibfield  {journal} {\bibinfo  {journal} {Physics
  of Fluids}\ }\textbf {\bibinfo {volume} {32}},\ \bibinfo {pages} {091903}
  (\bibinfo {year} {2020})}\BibitemShut {NoStop}%
\bibitem [{\citenamefont {Panda}\ \emph {et~al.}(2022)\citenamefont {Panda},
  \citenamefont {Sharma},\ and\ \citenamefont {Kumar}}]{26panda2022}%
  \BibitemOpen
  \bibfield  {author} {\bibinfo {author} {\bibfnamefont {M.~K.}\ \bibnamefont
  {Panda}}, \bibinfo {author} {\bibfnamefont {P.}~\bibnamefont {Sharma}}, \
  and\ \bibinfo {author} {\bibfnamefont {S.}~\bibnamefont {Kumar}},\
  }\href@noop {} {\bibfield  {journal} {\bibinfo  {journal} {Physics of
  Fluids}\ }\textbf {\bibinfo {volume} {34}},\ \bibinfo {pages} {024108}
  (\bibinfo {year} {2022})}\BibitemShut {NoStop}%
\bibitem [{\citenamefont {Panda}\ and\ \citenamefont
  {Rajput}(2023)}]{27rajput2023}%
  \BibitemOpen
  \bibfield  {author} {\bibinfo {author} {\bibfnamefont {M.~K.}\ \bibnamefont
  {Panda}}\ and\ \bibinfo {author} {\bibfnamefont {S.~K.}\ \bibnamefont
  {Rajput}},\ }\href@noop {} {\bibfield  {journal} {\bibinfo  {journal}
  {Physics of Fluids}\ }\textbf {\bibinfo {volume} {35}} (\bibinfo {year}
  {2023})}\BibitemShut {NoStop}%
\bibitem [{\citenamefont {Hill}\ and\ \citenamefont
  {H{\"a}der}(1997)}]{10hill1997}%
  \BibitemOpen
  \bibfield  {author} {\bibinfo {author} {\bibfnamefont {N.~A.}\ \bibnamefont
  {Hill}}\ and\ \bibinfo {author} {\bibfnamefont {D.-P.}\ \bibnamefont
  {H{\"a}der}},\ }\href@noop {} {\bibfield  {journal} {\bibinfo  {journal}
  {Journal of theoretical biology}\ }\textbf {\bibinfo {volume} {186}},\
  \bibinfo {pages} {503} (\bibinfo {year} {1997})}\BibitemShut {NoStop}%
\bibitem [{\citenamefont {Cash}\ and\ \citenamefont
  {Moore}(1980)}]{11cash1980}%
  \BibitemOpen
  \bibfield  {author} {\bibinfo {author} {\bibfnamefont {J.~R.}\ \bibnamefont
  {Cash}}\ and\ \bibinfo {author} {\bibfnamefont {D.~R.}\ \bibnamefont
  {Moore}},\ }\href@noop {} {\bibfield  {journal} {\bibinfo  {journal} {BIT
  Numerical Mathematics}\ }\textbf {\bibinfo {volume} {20}},\ \bibinfo {pages}
  {44} (\bibinfo {year} {1980})}\BibitemShut {NoStop}%
\bibitem [{\citenamefont {Bees}(2020)}]{15bees2020}%
  \BibitemOpen
  \bibfield  {author} {\bibinfo {author} {\bibfnamefont {M.~A.}\ \bibnamefont
  {Bees}},\ }\href@noop {} {\bibfield  {journal} {\bibinfo  {journal} {Annual
  Review of Fluid Mechanics}\ }\textbf {\bibinfo {volume} {52}},\ \bibinfo
  {pages} {449} (\bibinfo {year} {2020})}\BibitemShut {NoStop}%
\bibitem [{\citenamefont {Javadi}\ \emph {et~al.}(2020)\citenamefont {Javadi},
  \citenamefont {Arrieta}, \citenamefont {Tuval},\ and\ \citenamefont
  {Polin}}]{16javadi2020}%
  \BibitemOpen
  \bibfield  {author} {\bibinfo {author} {\bibfnamefont {A.}~\bibnamefont
  {Javadi}}, \bibinfo {author} {\bibfnamefont {J.}~\bibnamefont {Arrieta}},
  \bibinfo {author} {\bibfnamefont {I.}~\bibnamefont {Tuval}}, \ and\ \bibinfo
  {author} {\bibfnamefont {M.}~\bibnamefont {Polin}},\ }\href@noop {}
  {\bibfield  {journal} {\bibinfo  {journal} {Philosophical Transactions of the
  Royal Society A}\ }\textbf {\bibinfo {volume} {378}},\ \bibinfo {pages}
  {20190523} (\bibinfo {year} {2020})}\BibitemShut {NoStop}%
\bibitem [{\citenamefont {Ghorai}(1997)}]{19ghorai1997bioconvection}%
  \BibitemOpen
  \bibfield  {author} {\bibinfo {author} {\bibfnamefont {S.}~\bibnamefont
  {Ghorai}},\ }\emph {\bibinfo {title} {Bioconvection and plumes.}},\
  \href@noop {} {Ph.D. thesis},\ \bibinfo  {school} {University of Leeds}
  (\bibinfo {year} {1997})\BibitemShut {NoStop}%
\bibitem [{\citenamefont {Woods}(1954)}]{20woods1954note}%
  \BibitemOpen
  \bibfield  {author} {\bibinfo {author} {\bibfnamefont {L.~C.}\ \bibnamefont
  {Woods}},\ }\href@noop {} {\bibfield  {journal} {\bibinfo  {journal}
  {Aeronautical Quarterly}\ }\textbf {\bibinfo {volume} {5}},\ \bibinfo {pages}
  {176} (\bibinfo {year} {1954})}\BibitemShut {NoStop}%
\end{thebibliography}%
	
\end{document}